\documentclass{aa}  
\usepackage{graphicx}
\usepackage{amsmath}
\usepackage{amsfonts}
\usepackage{amssymb}
\usepackage{gensymb}
\usepackage{textcomp}
\usepackage[varg]{txfonts}
\usepackage{natbib}
\usepackage{epstopdf}
\bibpunct{(}{)}{;}{a}{}{,}
\begin{document}  


\title{Energetics of magnetic transients in a solar active region plage}

   \author{Authors}
   \author{L.~P.~Chitta\inst{1}, A.~R.~C.~Sukarmadji\inst{2,1}, L.~Rouppe van der Voort\inst{3,4}, \and H.~Peter\inst{1}}

   \institute{Max-Planck-Institut f\"ur Sonnensystemforschung, Justus-von-Liebig-Weg 3, 37077 G\"ottingen, Germany\\
              \email{chitta@mps.mpg.de} 
              \and
              University of St Andrews, St Andrews, Fife KY16 9SS, UK
              \and 
              Rosseland Centre for Solar Physics, University of Oslo, P.O. Box 1029 Blindern, N-0315 Oslo, Norway
              \and
              Institute of Theoretical Astrophysics, University of Oslo, P.O. Box 1029 Blindern, N-0315 Oslo, Norway
              }

   \date{Received 31 October 2018 / Accepted 5 February 2019}

\abstract
{Densely packed coronal loops are rooted in photospheric plages in the vicinity of active regions on the Sun. The photospheric magnetic features underlying these plage areas are patches of mostly unidirectional magnetic field extending several arcsec on the solar surface.}
%
{We aim to explore the transient nature of the magnetic field, its mixed-polarity characteristics, and the associated energetics in the active region plage using high spatial resolution observations and numerical simulations.}
%
{We used photospheric Fe\,{\sc i}\,6173\,\AA\ spectropolarimetric observations of a decaying active region obtained from the Swedish 1-m Solar Telescope (SST). These data were inverted to retrieve the photospheric magnetic field underlying the plage  as identified in the extreme-ultraviolet emission maps obtained from the Atmospheric Imaging Assembly (AIA) on board the Solar Dynamics Observatory (SDO). To obtain better insight into the  evolution of extended unidirectional magnetic field patches on the Sun, we performed 3D radiation magnetohydrodynamic simulations of magnetoconvection using the \texttt{MURaM} code.}
%
{The observations show transient magnetic flux emergence and cancellation events within the extended predominantly unipolar patch on timescales of a few 100\,s and on spatial scales comparable to granules. These transient events occur at the footpoints of active region plage loops. In one case the coronal response at the footpoints of these loops is clearly associated with the underlying transient. The numerical simulations also reveal similar magnetic flux emergence and cancellation events that extend to even smaller spatial and temporal scales. Individual simulated transient events transfer an energy flux in excess of 1\,MW\,m$^{-2}$ through the photosphere.}
%
{We suggest that the magnetic transients could play an important role in the energetics of active region plage. Both in observations and simulations, the opposite-polarity magnetic field brought up by transient flux emergence cancels with the surrounding plage field.  Magnetic reconnection associated with such transient events likely conduits magnetic energy to power the overlying chromosphere and coronal loops.}

   \titlerunning{Energetics of magnetic transients in a solar active region plage}
   \authorrunning{L. P. Chitta et al.}
   \maketitle

\section{Introduction} \label{sec:intro}

The energization of the upper atmosphere of cool stars eludes comprehensive understanding. To differentiate heating mechanisms, the observational consequences of the spatial distribution of the energy input play a key role when the Sun is studied. The concentrated energy input at the apex of coronal loops leads to a characteristic temperature (and emission structure) that is
different from footpoint-concentrated heating \citep[][]{1998Natur.393..545P}. Recently, a consensus has been reached that many, maybe most, coronal structures are heated predominantly near their footpoints \citep[][]{2007ApJ...659.1673A}, with notable exceptions, for example, for large flares that are heated from the top through reconnection \citep[e.g.,][]{1994Natur.371..495M}. 
In line with the scenario of footpoint heating, photospheric observations at the base of some active region coronal loops highlight the presence of mixed-polarity magnetic field patches that exhibit flux cancellation, a process through which energy can be supplied to heat the overlying corona \citep[e.g.,][]{2017ApJS..229....4C,2018A&A...615L...9C}. Two-dimensional (2D) and 3D magnetic reconnection models predict that such magnetic flux cancellation events at the solar surface, which would of course be associated with prior flux emergence, would transport energy to the loops in the upper solar atmosphere \citep[][]{2018ApJ...862L..24P,2019ApJ...872...32S}. 

\begin{figure*}
\begin{center}
\includegraphics[width=\textwidth]{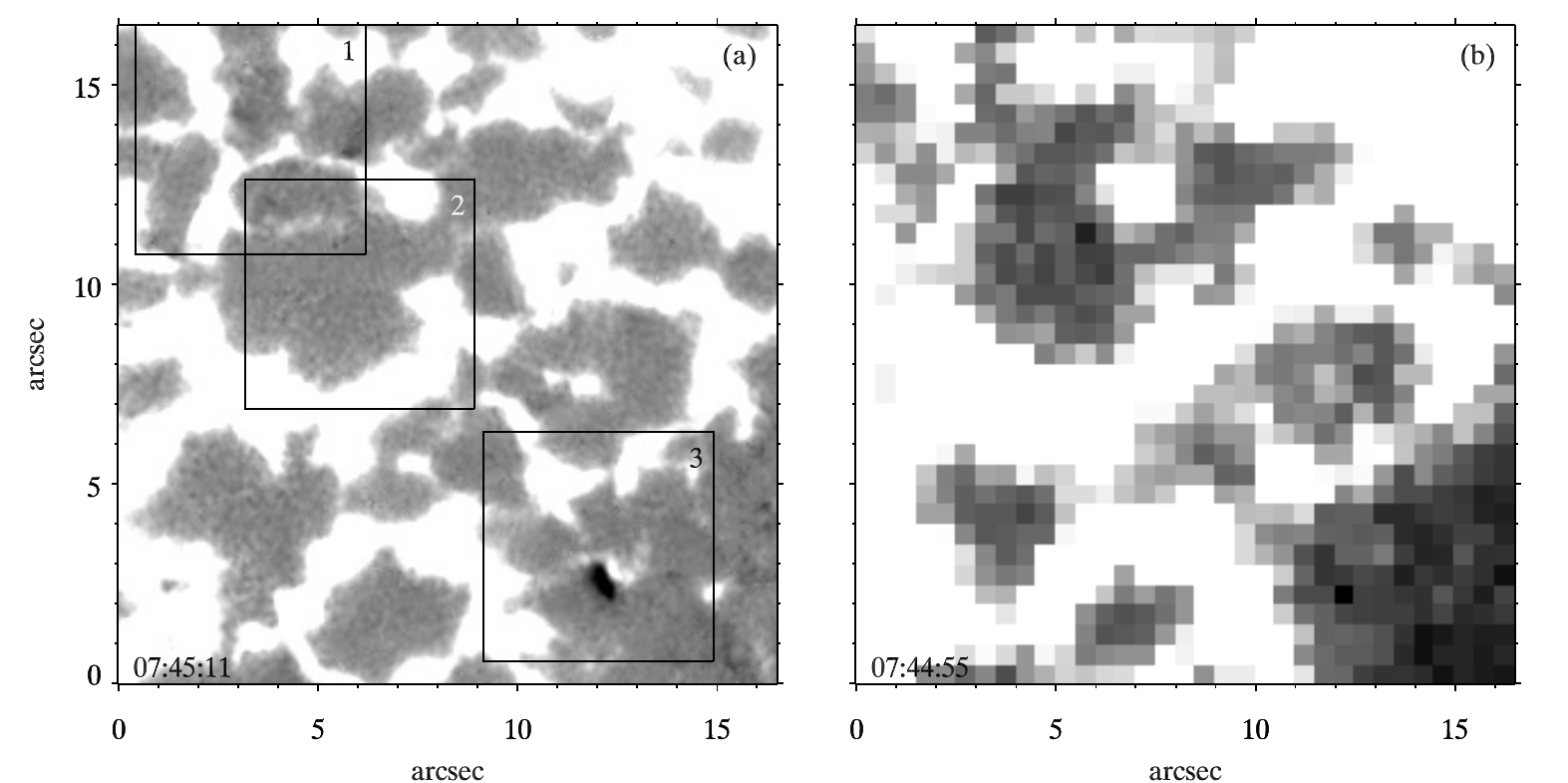}
\caption{Observations of the magnetic transients in a plage region. Details of the photospheric magnetic field in a decaying active region outlined by the black box in Fig.\,\ref{fig:sdoc}(c) are shown. Panel a: Line-of-sight magnetic field map obtained from SST observations on 26 June 2015 07:45\,UT. The three boxes (numbered 1--3; each covering an area of $5.76\arcsec\times5.76\arcsec$) enclose regions exhibiting transient magnetic flux emergence and cancellation at different instances during the evolution of a predominantly positive-polarity magnetic field patch in that active region (box 3 highlights one such transient).  Panel b: Line-of-sight magnetic field map obtained from the SDO/HMI displayed for comparison. Both maps are saturated at $\pm$100\,G. See Sect.\,\ref{sec:obs} for details.}
\label{fig:sstc}
\end{center}
\end{figure*}

More generally, however, at moderate spatial and temporal resolution, active regions are apparently covered by extended plage areas composed of moderately inclined unidirectional magnetic field patches \citep[e.g.,][]{2003ApJ...590..502D,2003ApJ...590..547A,2010ApJ...720.1380B,2010A&A...524A...3N,2015A&A...576A..27B}. Some of these unipolar plage areas are associated with so-called active region moss that is also considered to be heated from above. Moss is observed as a low-lying emission in the extreme-ultraviolet (EUV) at the footpoints of hot X-ray solar coronal loops and is structured on spatial scales of a megameter. It exhibits brightness variations on timescales on the order of 10\,s. Traditionally, this moss emission has been considered to be caused by heat conduction from the overlying hot loops \citep[][]{1999SoPh..190..409B}. Some recent observations suggest that nonthermal particles produced in coronal nanoflares are responsible for the observed high variability in the moss \citep[][]{2014Sci...346B.315T}. Common to these interpretations is that the energy source driving the heating and dynamics of the moss is localized in the coronal parts of the hot loops, for instance, as a result of nanoflares in the upper parts of the loops. Alternatively, it is also possible that the energy transport from highly dynamic footpoint reconnection from flux cancellation events is responsible for the high variability seen in the moss at the footpoints of hot coronal loops \citep[][]{2018A&A...615L...9C}. Central to understanding these details is the question of energy transport into the corona and the inherent role of magnetic field dynamics in the process. This question pertains also to plage areas where high-rising densely packed EUV loops are rooted (i.e., without hot X-ray loops and without low-lying moss-like EUV emission).

Although plage areas are predominantly unipolar, continued action of convection persists in these extended unidirectional magnetic field regions \citep[][]{2010A&A...524A...3N}. Occasionally, emergence of transient horizontal magnetic features driven by the upward-convective motions \citep[][]{2008ApJ...679L..57I} is also observed in plage areas, which results in a mixed-polarity magnetic field \citep[][]{2008A&A...481L..25I,2009A&A...495..607I}. However, the widespread role of these magnetic transients in plages and in the dynamics and energetics of both moss (i.e. with hot loops) and non-moss coronal loops is not clear. 

To discern the properties and energetics of such emerging transient horizontal magnetic features driven by magnetoconvection, we compared high-resolution observations of the photospheric magnetic field of a plage area underlying the footpoints of densely packed coronal loops in a decaying active region, with realistic 3D radiation magnetohydrodynamic (MHD) simulations representing the loop footpoints. Our observations show granular-scale transient magnetic flux emergence events at these loop footpoints. They evolve on timescales of about 5\,minutes, which is comparable to the lifetime of granules. The numerical simulations reproduce these emerging features, which persist over the horizontal extent of the simulation domain. They transfer an energy flux in excess of 1\,MW\,m$^{-2}$ through the solar photosphere. Provided that these simulations are a true representation of the real Sun, such an energy flux is in principle sufficient to power the emission from the overlying coronal loops. Our results indicate the important role of magnetic transients in the energetics of extended unipolar active region plage areas.

\begin{figure*}
\begin{center}
\includegraphics[width=\textwidth]{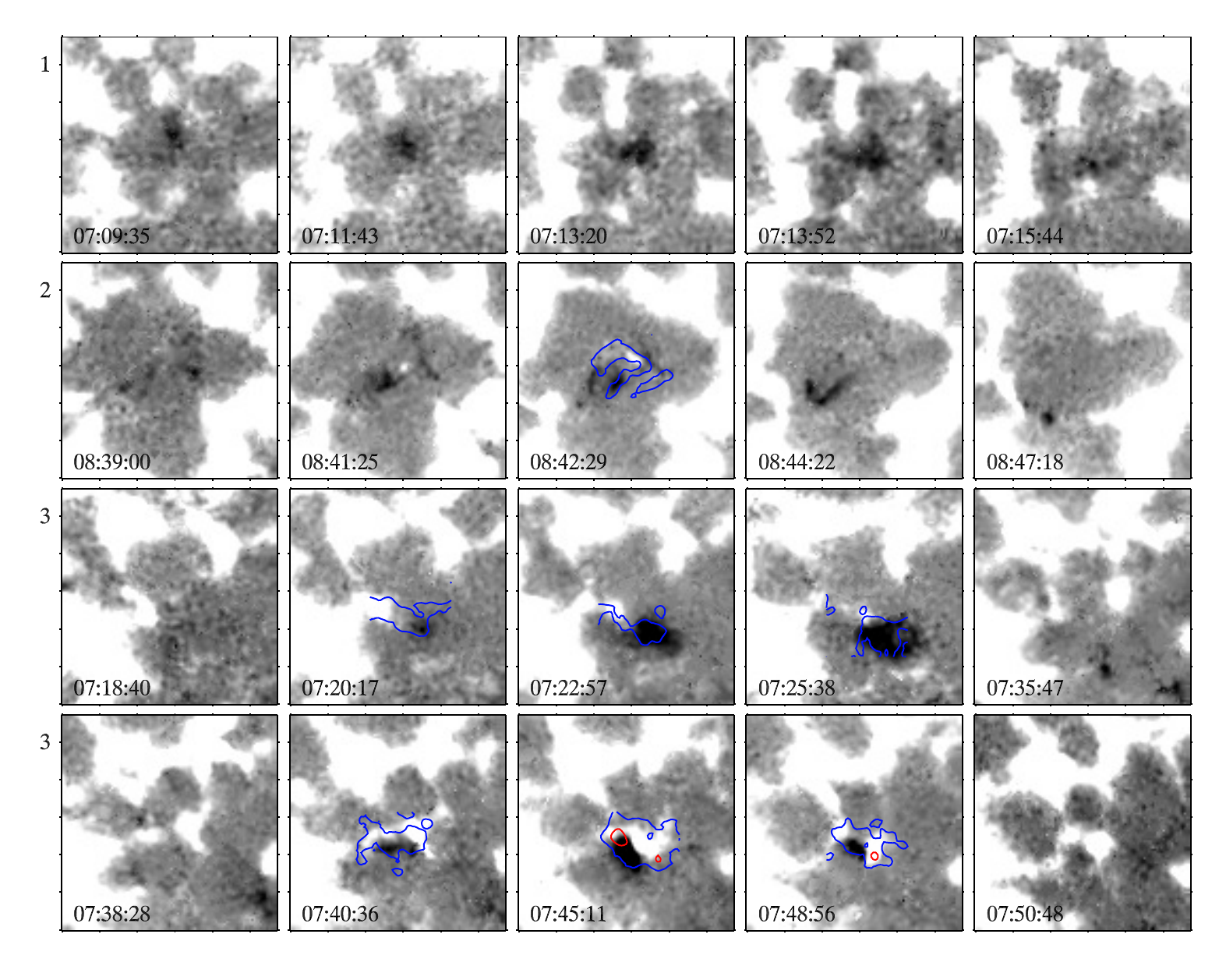}
\caption{Evolution of the magnetic transients in a plage region. The region covered by the panels in each row shows the SST line-of-sight magnetic field map (saturated at $\pm$50\,G) and is marked with the respective numbered box in Fig.\,\ref{fig:sstc}a. The rows display sequences of transient flux emergence and cancellation. The contours are for the horizontal component of the magnetic field (blue: 150\,G; red: 300\,G; the contours are restricted to a small rectangular patch that covers the transient). The bottom two rows show region 3 in Fig.\,\ref{fig:sstc}. The left-most column corresponds to the state prior to the emergence of transients (except for region 1, where the emergence had already begun at the start of SST observations). The right-most column shows the state after the cancellation of transient events. See Sect.\,\ref{sec:obs} for details.}
\label{fig:sstem}
\end{center}
\end{figure*}

\begin{figure}
\begin{center}
\includegraphics[width=0.5\textwidth]{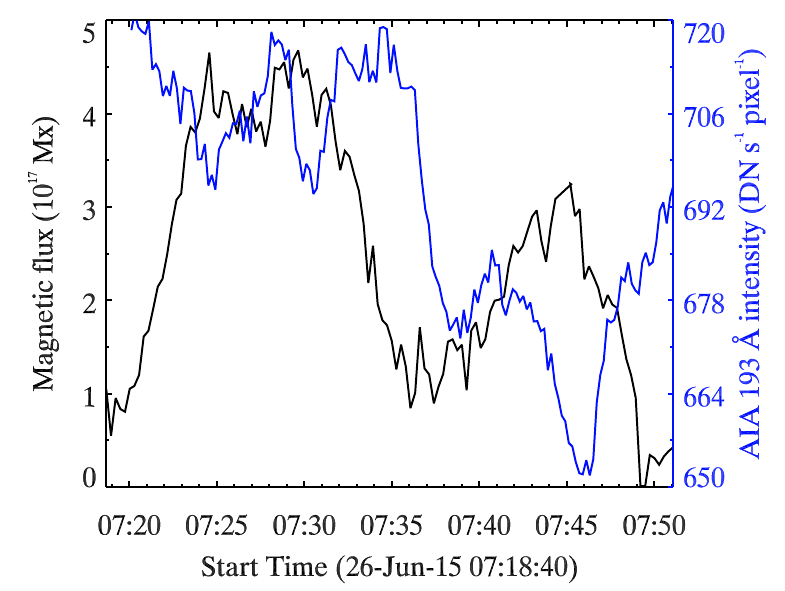}
\caption{Signatures of a magnetic transient in the photosphere and the corona. The black curve shows the photospheric magnetic flux integrated over the area that is covered by the minor negative-polarity features as a function of time from region 3 in Fig.\,\ref{fig:sstc}(a) from SST observations (see also Fig.\,\ref{fig:sstem}). To avoid noise, only pixels with a magnetic flux density above 10\,G\ are considered for the integration. The blue curve shows the coronal EUV emission recorded by the AIA 193\,\AA\ channel from the blue box in Fig.\,\ref{fig:sdoc}(d) overlying SST region 3. See Sect.\,\ref{sec:obs} for details.}
\label{fig:lcurves}
\end{center}
\end{figure}

\section{Observations of magnetic transients in a plage} \label{sec:obs}

To examine the evolution of predominantly unidirectional magnetic field patches on the solar photosphere (see the Appendix\,\ref{sec:uni} for details), we considered observations of the decaying active region from the Swedish 1-m Solar Telescope \citep[SST;][]{2003SPIE.4853..341S}, covering the area marked by the black box in Fig.\,\ref{fig:sdoc}c. The observational target of SST is a plage region underlying non-moss coronal region. Several faint coronal fan loops are observed to be rooted in that plage area (see Fig.\,\ref{fig:sdoc}d). In particular, we used spectropolarimetric observations of the Fe\,{\sc i}\,6173\,\AA\ line using the CRisp Imaging SpectroPolarimeter \citep[][]{2008ApJ...689L..69S} on the SST to determine the magnetic field properties in that region. These SST/CRISP data were recorded on 26 June 2015 between 07:09\,UT\ and 09:40\,UT. The full Stokes profile of the Fe\,{\sc i} line is obtained at eight positions between $\pm135$\,m\AA\ centered on 6173\,\AA, including a far red-wing position at $+350$\,m\AA. These data were processed following the standard CRISPRED pipeline \citep[][]{2015A&A...573A..40D}, which includes multi-object multi-frame blind deconvolution image restoration \citep[][]{2005SoPh..228..191V}. The restored images have a pixel scale of 0.057\arcsec\ and a cadence of 16\,s. The spatial resolution of the SST is five times as high as that of the Helioseismic and Magnetic Imager (HMI) on board the SDO. 

\begin{figure}
\begin{center}
\includegraphics[width=0.49\textwidth]{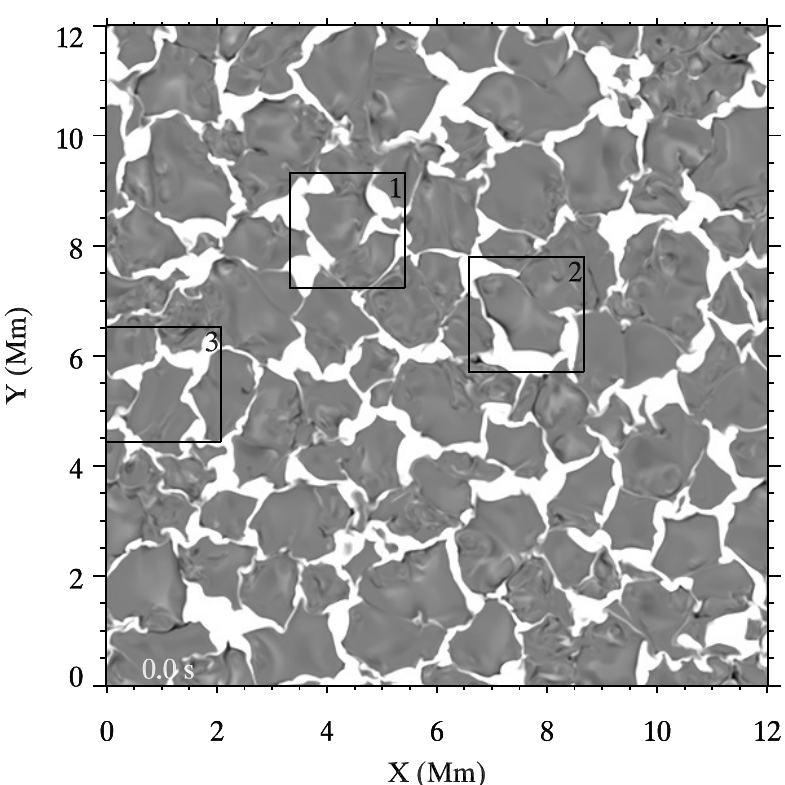}
\caption{Simulation of a plage region. The map is a 2D snapshot from a 3D magnetoconvection simulation showing the vertical component of the magnetic field saturated at $\pm$700\,G. This snapshot roughly corresponds to the average $\tau_{500}=1$ surface. Similar to Fig.\,\ref{fig:sstc}, the three boxes (numbered 1--3) enclose regions exhibiting transient magnetic flux emergence (cf. Fig.\,\ref{fig:emer}). Each box covers an area of $2\,\text{Mm}\times2\,\text{Mm}$ in the simulation domain. See Sect.\,\ref{sec:sim} for details.}
\label{fig:sim}
\end{center}
\end{figure}

These spectropolarimetric observations were inverted using SPINOR \citep[][]{2000A&A...358.1109F} to obtain the magnetic field vector. The average magnetic field strength in this predominantly unipolar region is about 250\,G. The line-of-sight component of the derived magnetic field from SST observations that are cospatial with the black box in Fig.\,\ref{fig:sdoc}c is displayed in the left panel of Fig.\,\ref{fig:sstc}. When observed at a spatial resolution of about 0.2\arcsec, the strong magnetic field is structured in intergranular lanes \citep[see][for general properties of such magnetic features]{2005A&A...435..327R}. While these SST observations also show a predominantly positive-polarity magnetic field in that region, we note several instances of mixed-polarity magnetic features that result from flux emergence within the granules. One such case is outlined by region 3. For comparison, the map of the line-of-sight magnetic field from the same region as observed with the SDO/HMI is displayed in Fig.\,\ref{fig:sstc}(b). Because its spatial resolution is modest, the SDO/HMI recorded only a coarse large-scale pattern of the dominant positive-polarity magnetic field in this plage region and does not show any signature of the small-scale minor negative-polarity field.  

The evolution of two apparent emerging features in region 3 is shown in the lower two rows of Fig.\,\ref{fig:sstem}. In the third row, a patch of minor negative-polarity magnetic field appears at 07:20\,UT. Initially, this element is connected to its conjugate positive-polarity magnetic field by a coherent horizontal magnetic field patch (blue contours)\footnote{The positive-polarity magnetic field patch associated with the transient event displayed in the third row emerges into the existing larger feature of the same polarity. For this reason, the horizontal magnetic field contours are open at the intersection of the two features.}. As the emergence proceeds, the negative-polarity element grows in size and recedes from its counterpart, at which time the horizontal magnetic field becomes more discrete. Finally, the negative-polarity magnetic element cancels with the surrounding opposite polarity field. After about 5\,minutes, a new transient event with the emergence of negative-polarity magnetic field is observed in the same region (fourth row). Unlike the previous case, the two polarities here do not move farther apart. They instead exhibit a twisting motion, with an apparent rotation of the negative-polarity feature with respect to its positive-polarity counterpart. A sample of other flux emergence and cancellation events from regions 1 and 2 is shown in the top two rows in Fig.\,\ref{fig:sstem}. These flux emergence events, which bring opposite-polarity magnetic features to the photosphere, lasted for about 5 to 10\,minutes, which is comparable to the lifetime of granules. The overall extended patch lasted for several hours. For this reason, we call these emerging and cancelling events magnetic transients. These events are similar to those observed in the quiet-Sun regions \citep[e.g.,][]{2009ApJ...700.1391M}.  

These magnetic transients are observed at the footpoints of coronal loops (cf. Fig.\,\ref{fig:sdoc}d). The opposite-polarity magnetic field brought up through emergence in plage regions could interact with the existing overlying field and supply energy to the coronal loops. In fact, such embedded small-scale opposite polarity magnetic patches can trigger magnetic reconnection in the  lower solar atmosphere \citep[e.g.,][]{2017A&A...605A..49C,2018A&A...617A.128S}. 

To illustrate the possible role of magnetic transients in coronal dynamics, we considered the example of two such events seen in region 3 (cf. Fig.\,\ref{fig:sstem}). We plot the integrated magnetic flux of the minor negative-polarity magnetic field in that region, which displays two distinct phases corresponding to the two events (black curve in Fig.\,\ref{fig:lcurves}). During these phases, magnetic flux associated with the negative-polarity field increases and decreases. In the first of two phases between 07:20\,UT and 07:25\,UT, the flux grows from $1\times10^{17}$\,Mx to  $4.5\times10^{17}$\,Mx in 300\,s, which amounts to a flux emergence rate on the order of $10^{15}$\,Mx\,s$^{-1}$. It also exhibits flux cancellation at the same rate from 07:30\,UT to 07:35\,UT. During the second phase, the flux emergence rate is slightly lower at $6\times10^{14}$\,Mx\,s$^{-1}$. The emission from the coronal loop footpoints overlying these transients in region 3 is plotted as a blue curve (cf. the blue box in Fig.\,\ref{fig:sdoc}d for the exact location of the loop footpoints). The coronal emission closely follows the evolution of the underlying magnetic transients with an apparent lag of about 5\,minutes. A rapid drop in the coronal emission soon after the cancellation of the negative-polarity magnetic field at the end of the first event around 07:35\,UT is very clear. This demonstrates the scenario in which the small-scale transient events at the photosphere trigger a response in the corona. Flux emergence and cancellation are widely associated with magnetic reconnection, which could also be the process responsible for the modulation of coronal emission in this case.

\begin{figure*}
\begin{center}
\includegraphics[width=\textwidth]{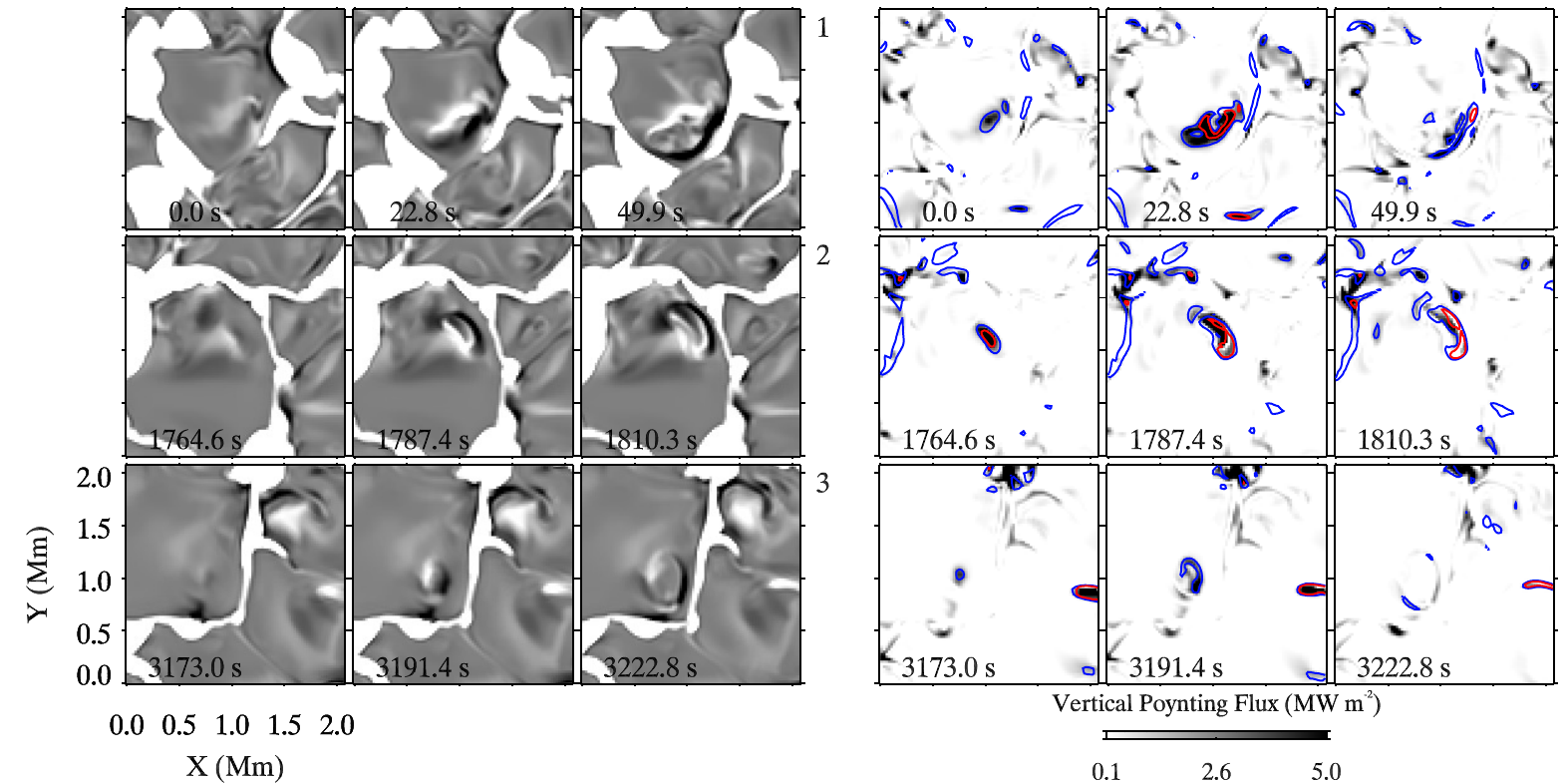}
\caption{Simulated magnetic transients. The left panels show events of transient flux emergence in a plage simulation. Each row (numbered 1--3) tracks the evolution of flux emergence in granules from the respective areas marked in Fig.\,\ref{fig:sim}. The vertical component of the magnetic field is saturated at $\pm300$\,G. The right panels display the spatial association of the photospheric horizontal magnetic field and the vertical component of the Poynting flux. These snapshots correspond to those displayed in the left panels. Shown in grayscale is the vertical component of the Poynting flux. The blue and red contours outline regions with a horizontal magnetic strength of 400\,G\ and 600\,G, respectively. These snapshots roughly correspond to the average $\tau_{500}=1$ surface. See Sect.\,\ref{sec:sim} for details.}
\label{fig:emer}
\end{center}
\end{figure*} 

However, given the diffused and large-scale structuring (about 10\arcsec) of the loop footpoints in the corona, it is difficult to associate every photospheric magnetic transient with a dynamic signature in the corona. Furthermore, the energy flux transferred by these events through the solar surface is difficult to estimate properly from observations alone. Therefore, numerical experiments are imperative to study the energetics of magnetic transients discussed here.

Although this AR was decaying, granular-scale flux emergence was still ongoing. This is due to the magnetoconvection that operates on the predominantly positive-polarity field. Thus even in the almost unipolar magnetized region, convection will bring opposite polarities to the surface. We studied this process through 3D MHD simulations.

\section{Simulations of magnetic transients in a plage}  \label{sec:sim}

We used high-resolution simulations to verify the presence of magnetic transients in regions with a predominantly unidirectional magnetic field. These simulations will also provide better information on photospheric flow and magnetic field properties to estimate the energetics of magnetic transients. To this end, we employed 3D radiation MHD simulations using the \texttt{MURaM} code \citep[][]{2005A&A...429..335V}. This code has been extensively used to simulate magnetoconvection near the solar surface covering a variety of magnetic field configurations ranging from the quiet Sun to pores to sunspots \citep[e.g.,][]{2006ApJ...641L..73S,2007A&A...465L..43V,2007A&A...474..261C,2014ApJ...785...90R}. Dynamic phenomena such as magnetic flux emergence \citep[e.g.,][]{2007A&A...467..703C,2009A&A...507..949T}, vortices, shocks, waves \citep[e.g.,][]{2012A&A...541A..68M,2013ApJ...776L...4S}, or the generation of MHD Poynting flux \citep[][]{2012ApJ...753L..22S} have also been widely studied  with \texttt{MURaM}.

To simulate the magnetoconvection at the footpoints of a coronal fan loop in a plage region, we focused on a shallow region near the solar surface that extends to 1.4\,Mm\ in height, where the top of the simulation domain reaches 600\,km\ above a layer of optical depth at 500\,nm\ of $\tau_{500}=1$. This vertical extent is sampled by a grid of 128 pixels. The simulation domain is periodic in both horizontal directions and covers an area of $12\,\text{Mm}\times12\,\text{Mm}$, with a mesh size of $512\times512$ pixels. This $12\,\text{Mm}\times12\,\text{Mm}$ horizontal extent of simulation is equivalent to the area covered by the black boxes in Fig.\,\ref{fig:sdoc} or the field of view shown in Fig.\,\ref{fig:sstc}. The bottom boundary is open for an in- and outflow of mass, and the convective flux vanishes at the top boundary. These simulations were run with non-gray radiative transfer including the effects of partial ionization. The domain was first evolved without a magnetic field until steady-state convection set in. Next, to mimic the SST observations of loop footpoints that contain an average magnetic field strength of about 250\,G\ with predominantly unidirectional magnetic field, we added a uniform magnetic field of 200\,G\ in the vertical direction to the simulation domain. At the top and bottom boundaries, the magnetic field was set to remain vertical. 

The convection advects and concentrates this uniform magnetic field in the intergranular lanes. A sample snapshot of this structured magnetic field of positive polarity in the computational domain from a height corresponding roughly to the average $\tau_{500}=1$ surface is shown in Fig.\,\ref{fig:sim}. The snapshot also displays patches of weaker magnetic field with a polarity opposite to that of the main polarity. These minor negative-polarity magnetic features arise from small-scale flux emergence events in the granules, which is a natural consequence of magnetoconvection. Three such events are displayed in the left panels of Fig.\,\ref{fig:emer}. They are similar to the observed events in that they have a lifetime on the order of 100\,s\ and are randomly distributed over the horizontal extent of the domain. Weaker transients with shorter timescales ($\sim$50\,s) are also revealed in the simulations, but their evolution is more difficult to follow. 

\begin{figure}
\begin{center}
\includegraphics[width=0.49\textwidth]{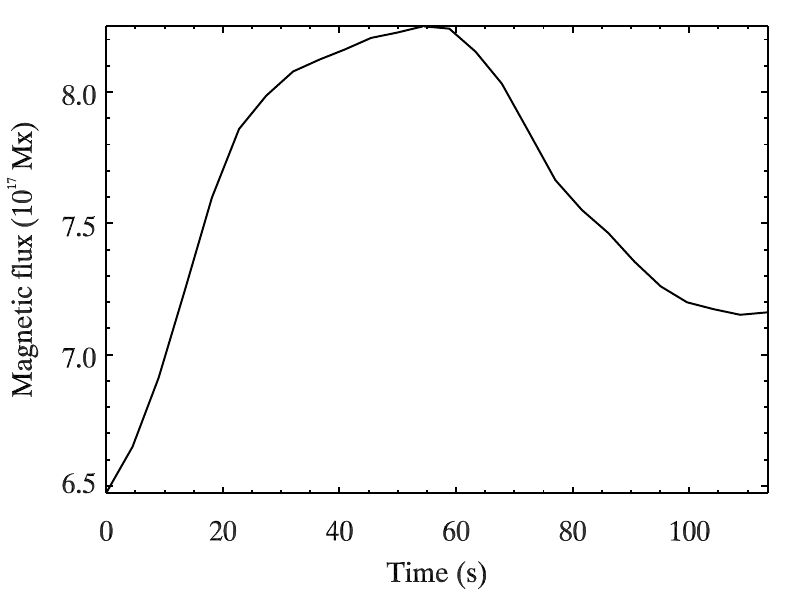}
\caption{Flux emergence and cancellation in simulations. The emergence and cancellation of minor negative-polarity magnetic field in a transient event from region 1 in Fig.\,\ref{fig:sim} is plotted as a function of time (cf. the top row in the left panel in Fig.\,\ref{fig:emer}). See Sect.\,\ref{sec:sim} for details.}
\label{fig:mflux}
\end{center}
\end{figure}

It is clear from these images that the flux cancellation follows the emergence as the minor opposite-polarity field cancels with the main polarity field. To demonstrate this flux cancellation, we considered the case of transient from region 1 in the simulations. The evolution of the integrated magnetic flux of minor negative polarity from this region is shown in Fig.\,\ref{fig:mflux}. Similar to the observational case in Fig.\,\ref{fig:lcurves}, the flux emergence and cancellation rates for this simulated transient are on the order of $10^{15}$\,Mx\,s$^{-1}$.

The interaction of upward and horizontal surface motions with the magnetic field generates Poynting flux, which can propagate into the upper atmosphere. The Poynting flux vector due to surface motions is given by $\boldsymbol{S}=\frac{1}{4\pi}\boldsymbol{B}\times(\boldsymbol{v}\times\boldsymbol{B})$, where $\boldsymbol{v}$ is the flow field and $\boldsymbol{B}$ is the magnetic field.\footnote{This description of Poynting flux ignores the contribution of currents. Currents will become important in the corona \citep[][]{2009PhDT.........1B}.} In particular, we are interested in the vertical component of the Poynting flux, $\boldsymbol{S}_z$, which is injected into the upper atmosphere. The quantity $\boldsymbol{S}_z$ has contributions from both the upward convective motions and the horizontal surface motions. We find that during the initial phases of the flux emergence, upward motions contribute to $\boldsymbol{S}_z$, and at later phases, when the emerged flux is advected toward intergranular lanes, horizontal surface motions dominate the energy flux. Given the transient nature of the events, both these contributions must be taken into account to compute $\boldsymbol{S}_z$. 

We show the association of Poynting flux with the transient emergences in the right panels of Fig.\,\ref{fig:emer}. The regions of higher $\boldsymbol{S}_z$ (negative grayscale maps) are related to stronger horizontal magnetic field strengths (blue and red contours). By comparing them with the events in the left panels of the figure, it is clear that this strong horizontal magnetic field is a result of flux emergence at the surface. This horizontal magnetic field is associated with the vertical Poynting flux. As the flux emergence proceeds above the surface, we tracked the transfer of $\boldsymbol{S}_z$ and its evolution in time (Fig.\,\ref{fig:pftra}; see Appendix\,\ref{sec:app} for details). These transient events inject an energy flux $\boldsymbol{S}_z$ of more than 1\,MW\,m$^{-2}$ through the solar surface. 

\begin{figure}
\begin{center}
\includegraphics[width=0.49\textwidth]{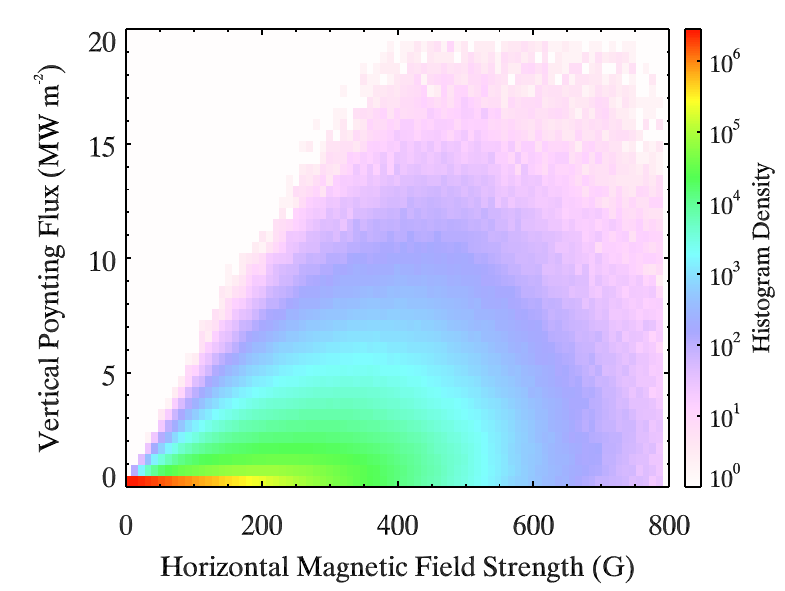}
\caption{Statistical association of photospheric horizontal magnetic field and the vertical component of the Poynting flux. The histogram displayed in logarithmic scaling is a joint probability density function between horizontal magnetic field strength and the vertical component of the Poynting flux at the photosphere, sampled for a period of 15\,minutes. The two quantities roughly correspond to the average $\tau_{500}=1$ surface. See Sect.\,\ref{sec:sim} for details.}
\label{fig:stat}
\end{center}
\end{figure}

The observational case presented in Fig.\,\ref{fig:lcurves} suggests that despite its small-scale nature, the transient event could trigger a coronal response. This requires some process by which the energy flux associated with the transient near the photosphere is carried upward. To this end, our simulations hint at a rapid reconfiguration of the overlying magnetic field during the transient flux emergence and cancellation (Fig.\,\ref{fig:flines}). We speculate that this reconfiguration is facilitated by magnetic reconnection above the solar surface, which enables the energy flux to reach the upper atmosphere. 

The maps of $\boldsymbol{S}_z$ provide a spatial association with the horizontal magnetic field. In the case of these transients, however, the horizontal magnetic field at the surface decreases while the flux continues to emerge. Afterward, $\boldsymbol{S}_z$ is generated by the horizontal surface motions of the pair of footpoints that were previously connected by the horizontal magnetic field. The general relation of $\boldsymbol{S}_z$ with the horizontal magnetic field is of interest because it can be used as a diagnostic with observations. In Fig.\,\ref{fig:stat} we plot the joint probability density function (PDF) between horizontal magnetic field strength and $\boldsymbol{S}_z$. This PDF was obtained from 200 simulation snapshots covering 15\,minutes of solar evolution to capture a variety of emerging events in their early phases\footnote{Magnetoconvection in the photosphere will sufficiently evolve during the period of 15\,minutes, which is long enough compared to the typical granular timescales of 5--8\,minutes.}. For field strengths above 200\,G,\ a broad tail of $\boldsymbol{S}_z$ with values in excess of 5\,MW\,m$^{-2}$ is observed. 

In active regions, the chromosphere typically requires an energy budget of 20\,kW\,m$^{-2}$, and in the corona, the requirement is 10\,kW\,m$^{-2}$ \citep[][]{1977ARA&A..15..363W}. This means that only a fraction of the energy flux carried by a magnetic transient may be required to power individual active region loops overlying plage areas, while most of it is expected to be dissipated deep in the solar atmosphere as a result of the small-scale nature of these events. The question remains what happens with the energy flux that is made available by transients for the whole plage region. Here we provide a preliminary but conservative estimate of the average energy flux associated with the magnetic transients in plage regions by considering the event from region 1 as a prototype  (cf. Figs.\,\ref{fig:sim} and \ref{fig:emer}). Based on the emergence and cancellation of the minor opposite-polarity magnetic field, the event lasted for about 100\,s (see Fig.\,\ref{fig:mflux}). During this period, the transient event possessed a total energy of $E=3\times10^{19}$\,J. This energy is about two orders of magnitude higher than a typical coronal nanoflare, which is estimated to release an energy of $10^{17}$\,J \citep[][]{1988ApJ...330..474P}. In our simulations there is at least one such transient event in an area of $2\,\text{Mm}\times2\,\text{Mm}$ over the course of 1\,hour of solar evolution. Thus the total number of events, $n$, is at least 36 during the time period, $t=3600$\,s, in these simulations that cover an area $A=12\,\text{Mm}\times12\,\text{Mm}$. From these values, we estimate a lower limit on the average energy flux associated with simulated magnetic transients, $nE/At$ of 2\,kW\,m$^{-2}$. A proper energy flux estimate requires a statistics from the evolution of individual transients, which is a study on its own. Nevertheless, our conservative lower limit of the average energy flux based on simulations suggests that the magnetic transients could play an important role in the energetics of an active region plage in general, supporting the conclusions of previous studies by \citet{2009A&A...495..607I} and \citet{2008ApJ...679L..57I}.

\section{Conclusions} \label{sec:conc}

We used the SST photospheric observations and \texttt{MURaM} 3D MHD simulations to study the evolution of the magnetic field in a plage area at the footpoints of a system of non-moss fan loops that we identified in SDO/AIA images. The small granular-scale flux emergence and cancellation events found both in  the observations and simulations suggest a very dynamic evolution of mostly unipolar magnetic plages on the Sun at small spatial scale. In one case, the emission from the footpoints of coronal loops closely followed the evolution of magnetic elements undergoing emergence and cancellation (cf. Fig.\,\ref{fig:lcurves}). These findings support the observational study of \citet{2009A&A...495..607I}. The authors suggested that magnetic transients in plage regions could contribute to chromospheric and coronal heating. Simulations show that magnetoconvection works on the (initially) vertical magnetic field and brings up small bipoles that appear as small patches of opposite magnetic polarity in magnetograms. These loop-like emerging events transfer an energy flux in excess of 1\,MW\,m$^{-2}$ through the photosphere, part of which may reach the upper atmosphere through reconnection between the emerging and the overlying magnetic field, as concluded by \citet{2008ApJ...679L..57I}. One way to transfer this energy flux is through MHD waves. Numerical models show that reconnection between the emerging and overlying magnetic field can generate high-frequency Alfv\'{e}n waves that propagate into the solar corona \citep[][]{2008ApJ...679L..57I}. 

Connecting individual flux emergence events to dynamic signatures in the upper atmosphere, however, remains a major observational challenge. One reason for this is the spatial resolution disparity between photospheric and coronal observations. Nevertheless, coronal observations indicate such a reconnection process with persistent subsonic upflows and possible unresolved high-speed upflows near the footpoints of coronal loops \citep[][]{2008ApJ...678L..67H}. Our results are also relevant for the studies of coronal-hole plumes. It has been suggested that these structures are formed through the cancellation of the minor opposite-polarity magnetic field with the surrounding unipolar magnetic patches \citep[e.g.,][]{2014ApJ...787..118R}. Overall, the magnetic transients in our simulation provide an additional basis to recent observational studies that found flux cancellation at the footpoints of different coronal loop systems \citep[][]{2017ApJS..229....4C,2018A&A...615L...9C}. The upcoming 4-m Daniel K. Inouye Solar Telescope \citep[][]{2014SPIE.9147E..07E} will be able to resolve the photosphere down to about 30\,km, which might reveal the persistent flux emergence that has been predicted by simulations. Such observations will provide better constraints on the role of small-scale flux emergence and cancellation transients in the heating of active region moss, and the corona in general.

\begin{acknowledgements}
We thank the anonymous referee for constructive comments that helped to improve the manuscript. L.P.C. received funding from the European Union's Horizon 2020 research and innovation program under the Marie Sk\l{}odowska-Curie grant agreement No.\,707837. A.R.C.S. acknowledges funding from the Max Planck Institute for Solar System Research for an internship. L.R.v.d.V. is supported by the Research Council of Norway, project number 250810, and through its Centres of Excellence scheme, project number 262622. The Swedish 1-m Solar Telescope is operated on the island of La Palma by the Institute for Solar Physics of Stockholm University in the Spanish Observatorio del Roque de los Muchachos of the Instituto de Astrof\'{i}sica de Canarias. The Institute for Solar Physics is supported by a grant for research infrastructures of national importance from the Swedish Research Council (registration number 2017-00625). SDO data are courtesy of NASA/SDO and the AIA and HMI science teams. The numerical simulations presented in this work have been performed on the supercomputers at the GWDG. Fig.\,\ref{fig:flines} is produced by VAPOR (www.vapor.ucar.edu). This research has made use of NASA's Astrophysics Data System.
\end{acknowledgements}


\begin{appendix}

\section{Active region plages on the Sun} \label{sec:uni}

\begin{figure*}
\begin{center}
\includegraphics[width=\textwidth]{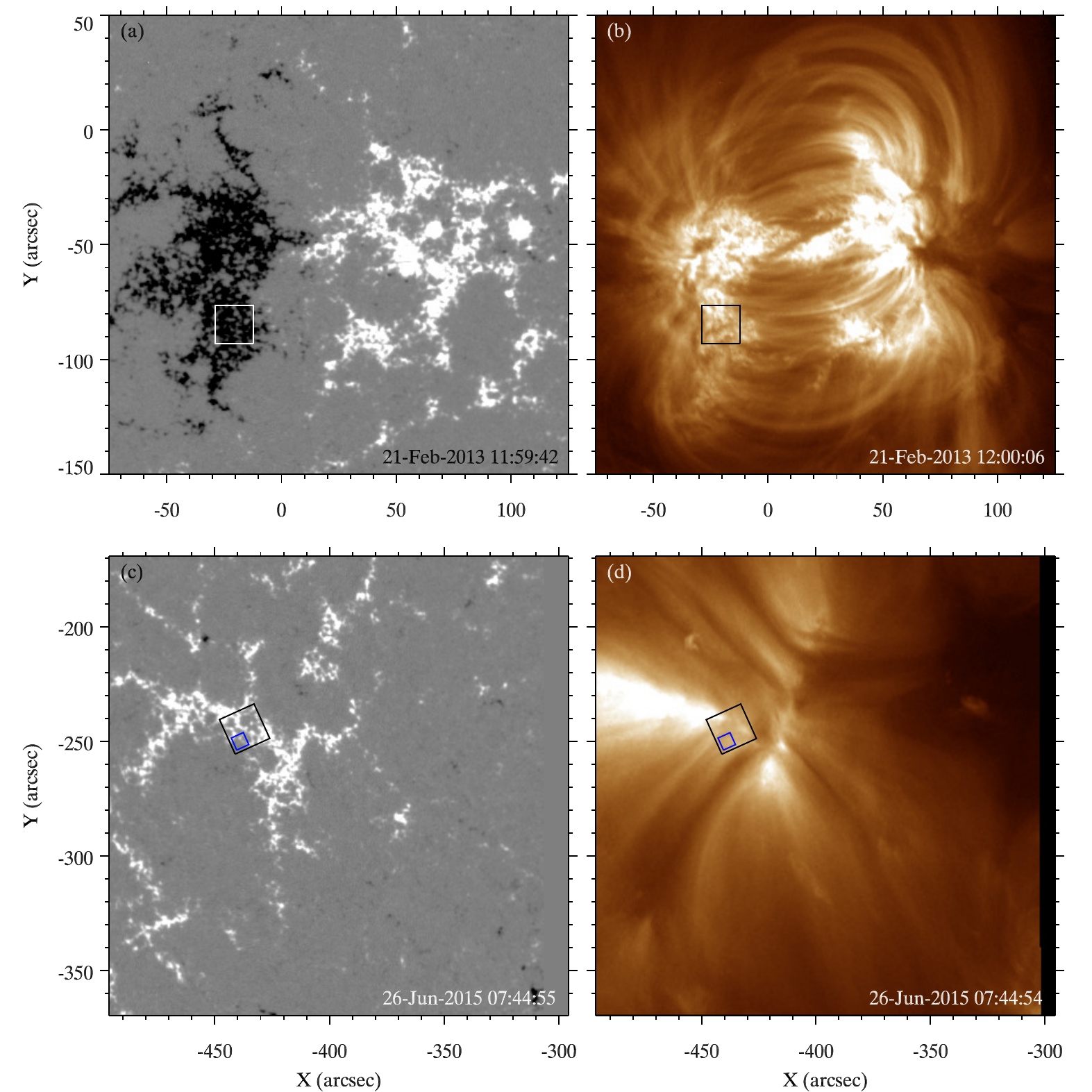}
\caption{Context maps showing the photospheric and coronal diagnostics of plage areas in active regions.  (a) The photospheric line-of-sight magnetic field map of an evolved active region obtained from SDO/HMI. The white and black shaded areas are the positive- and negative-polarity magnetic field regions, saturated at $\pm$300\,G. The white box covers a patch of the active region with a predominantly negative-polarity magnetic field in an area of $16.5\arcsec\times16.5\arcsec$. (b) The map of coronal emission of the active region recorded by the SDO/AIA 193\,\AA\ filter. The black box, which is cospatial with the white box in panel (a), identifies a coronal low-lying moss emission in this active region. Panels (c) and (d) are the same as panels (a) and (b), but plotted for a decaying active region (weaker than the above). The black boxes have the same size as in the top panels and cover a predominantly positive-polarity magnetic field patch in this active region. In the bottom panels the black box is the target region of the SST co-observations. The smaller blue box corresponds to region 3 in Fig.\,\ref{fig:sstc}. See Sect.\,\ref{sec:obs} for details.}
\label{fig:sdoc}
\end{center}
\end{figure*}

To lay out the context for the magnetic plages, we display snapshots of two active regions with a photospheric line-of-sight magnetic field (left panels) and the coronal emission near 1.5\,MK\ (right panels) in Fig.\,\ref{fig:sdoc}. The magnetic field maps are obtained from the Helioseismic and Magnetic Imager \citep[HMI;][]{2012SoPh..275..207S} on board the Solar Dynamics Observatory \citep[SDO;][]{2012SoPh..275....3P}. The coronal emission is recorded with the 193\,\AA\ filter on the Atmospheric Imaging Assembly \citep[AIA;][]{2012SoPh..275...17L} on board the SDO. 

The two active regions are at different stages of their evolution. These examples are selected to show the general characteristics of plages and the associated coronal emission (from the low-lying moss emission and the longer loops). The example in the top panels is an evolved region with a clear bipolar magnetic structure (left panel). The low-lying moss emission is evident from the AIA 193\,\AA\ snapshot (right panel). One such moss region is highlighted with a black box. The photospheric counterpart of this coronal moss is a densely structured plage of negative-polarity magnetic field, with an average magnetic flux density of about 300\,G. The smaller gray patches between the magnetic structures (in the boxed region) indicate photospheric granulation.  Thus the region would represent an extended patch of unidirectional magnetic field that forms the footpoint of the coronal moss. Similar features are found in the whole active region in both magnetic polarities. In the bottom panels, part of a decaying active region is displayed, where the positive polarity field is readily visible in the photosphere (bottom left panel). The associated trailing negative-polarity magnetic field in this active region is outside the displayed field of view. Unlike the low-lying moss features in Fig.\,\ref{fig:sdoc}b, the coronal emission in the decaying active region is structured as a longer fan-like loop system (black box in Fig.\,\ref{fig:sdoc}d). Again, the footpoints of these fan loops constitute an extended patch of unipolar magnetic field (positive polarity in this case) that is structured by granulation, with an average magnetic flux density of about 130\,G. At least when observed at a spatial resolution of about 1\arcsec\ with HMI, the footpoint patches in both active regions are predominantly unidirectional, without clear signs of a mixed-polarity magnetic field. 

\section{Transfer of Poynting flux through the solar photosphere} \label{sec:app}

\begin{figure}
\begin{center}
\includegraphics[width=0.49\textwidth]{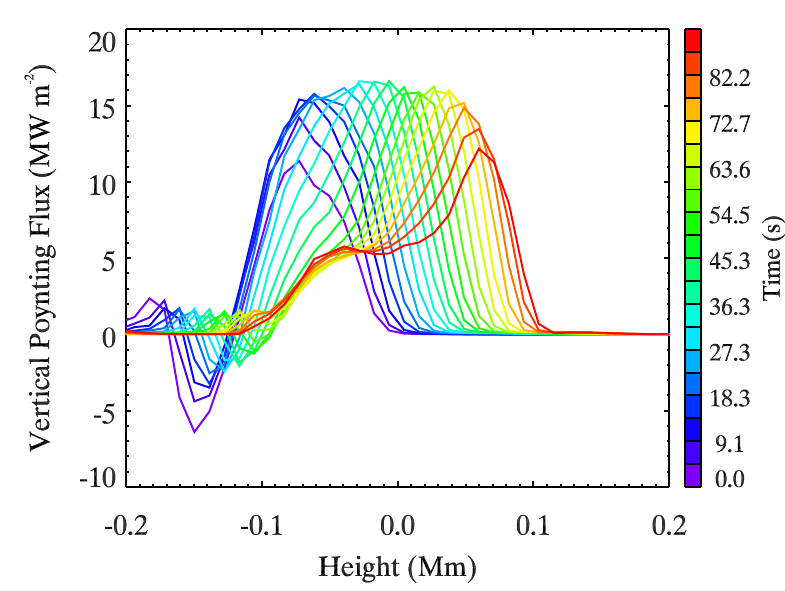}
\caption{Transfer of Poynting flux through the solar photosphere. Each curve represent the vertical component of the Poynting flux at a given location in a simulated flux emergence event, with in a height range of $\pm$0.2\,Mm near the solar surface. The colors denote the temporal evolution of the event. See Sect.\,\ref{sec:sim} for details.}
\label{fig:pftra}
\end{center}
\end{figure}

Magnetic flux emergence in our simulation transfers Poynting flux through the solar surface, which is sufficient to heat coronal fan loops in active region plage areas (cf. Sect.\,\ref{sec:sim}). These emerging events are transient in nature and generally carry an energy flux in excess of 1\,MW\,m$^{-2}$. The complete information on photospheric flows and magnetic field in 3D allows us to track the transfer of energy flux with height and time. This spatio-temporal profile of the Poynting flux, $\boldsymbol{S}$, for one such event is displayed in Fig.\,\ref{fig:pftra}. Here we trace a flux emergence event along a single vertical column in the 3D domain and plot the vertical component of the Poynting flux, $\boldsymbol{S}_z$, as a function of time. Shown here is only a phase of about 90\,s\ of the emergence event that lasted over 200\,s, which is comparable to the granular lifetime. In the initial phases of emergence (purple curves), the $\boldsymbol{S}_z$ that is generated below the surface increases in magnitude and at the same time is transferred across the surface. The extent of flux emergence increases in height with time, which is seen as broadening of the $\boldsymbol{S}_z$ profiles. This means that once the emergence process is triggered, the upward convective motions continually pushes more magnetic flux in the same region over an extended period. Overall, there is a net flux of magnetic energy that will be transported upward into the solar atmosphere. 

\begin{figure*}
\begin{center}
\includegraphics[width=\textwidth]{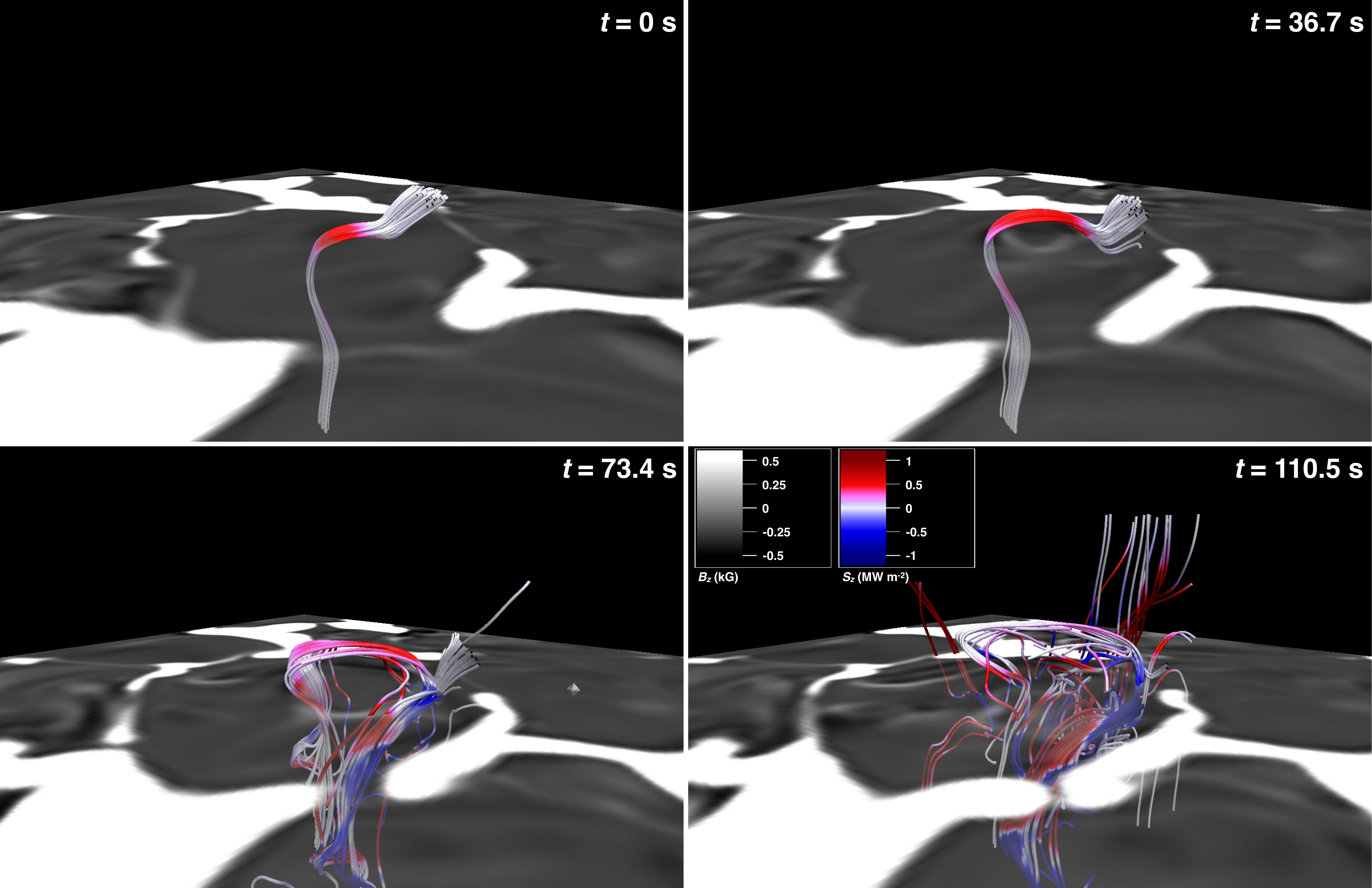}
\caption{Visualisation of a simulated magnetic flux transient. Displayed here are four instances of a small flux emergence event within a granule. The surface covers an area of $3\,\text{Mm}\times3\,\text{Mm}$. Each snapshot is about 37\,s\ apart. The grayscale image shows the vertical component of the magnetic field near the photosphere. The solid curves trace the magnetic field in the 3D domain. Each field line is color coded with the vertical component of the Poynting flux. See Sect.\,\ref{sec:sim} for details.}
\label{fig:flines}
\end{center}
\end{figure*}

The magnetic flux that emerged above the surface could interact with the overlying field through reconnection. This process will enable the transfer of energy flux from the emerging events to the solar atmosphere, which could then energize the coronal loops that are rooted in plage regions (cf. Fig.\,\ref{fig:lcurves}). Qualitative signatures of such reconnection are indeed seen in our simulation. In Fig.\,\ref{fig:flines} we show a 110\,s\ phase of a small-scale flux emergence and the subsequent evolution of field lines during the event within a granule\footnote{The event described in Fig.\,\ref{fig:flines} is different from the one discussed in Fig.\,\ref{fig:pftra}}. In the early phases of emergence, the field lines are more organized and show a tendency of looping back to the surface. The flux emergence becomes clear at later times with a small loop system. With the event progressing, the initially organized field lines become highly disorganized. This eruption-like behavior indicates a rapid reconfiguration of the magnetic field through reconnection, which could transfer energy flux to the upper atmosphere. Future observations are required to study the observational manifestation of these reconnection events and their impact on the energetics of the chromosphere and corona.  

\end{appendix}

\end{document}